\documentclass[review]{elsarticle}

\usepackage{lineno}
\biboptions{sort&compress}
\usepackage[colorlinks=true,linkcolor=blue,anchorcolor=blue,citecolor=blue]{hyperref}
\modulolinenumbers[5]
\usepackage{bm}
\usepackage{amsmath}

\begin{document}

\begin{frontmatter}

\title{Density-functional study of plutonium monoxide monohydride}
\author[mymainaddress]{Ruizhi Qiu}
\cortext[mycorrespondingauthor]{Corresponding author}
\ead{qiuruizhi@itp.ac.cn}
\author[mymainaddress]{Haiyan Lu}
\author[mymainaddress]{Bingyun Ao}
\author[mymainaddress]{Tao Tang}
\author[mymainaddress]{Piheng Chen}
\ead{chenph@live.cn}
\address[mymainaddress]{Science and Technology on Surface Physics and Chemistry Laboratory,
Mianyang 621908, Sichuan, P.R. China}
%\address[mysecondaryaddress]{China Academy Engineering Physics, Mianyang 621900, Sichuan, P.R. China}

\begin{abstract}
The structural, electronic, mechanical, 
optical, thermodynamic properties of plutonium monoxide monohydride (PuOH) are studied by density-functional calculations within the framework of LDA/GGA and LDA/GGA+$U$. 
From the total energy calculation, the lowest-energy crystal structure of PuOH is predicted to have space group $F{\bar4}3m$ (No.~216).
Within the LDA+$U$ framework, the calculated lattice parameter of $F\bar{4}3m$-PuOH is in good agreement with the experimental value and 
the corresponding ground state is predicted to be an antiferromagnetic charge-transfer insulator. 
Furthermore, we investigate the bonding character of PuOH by analyzing the electron structure and find that there are a stronger Pu-O bond and a weaker Pu-H bond.
The mechanical properties including the elastic constants, elastic moduli and Debye's temperature,
and the optical properties including the reflectivity and absorption coefficient are also calculated.
We then compute the phonon spectrum which verified the dynamical stability of $F\bar{4}3m$-PuOH.
Some thermodynamic quantities such as the specific heat are evaluated.
Finally we calculate the formation energy of PuOH,
and the reaction energies for the oxidation of PuOH and PuOH-coated Pu,
which are in reasonable agreement with the experimental values.
\end{abstract}

\end{frontmatter}

\linenumbers

\section{Introduction}

The corrosion of plutonium have been attracting considerable interest in surface chemistry, 
corrosion chemistry and condensed matter physics~\cite{Haschke2000,Haschke2000b,Butterfield2004,Haschke2001,Wang2012,Nelson2013,Nelson2015,Huda2005,Taylor2014}. 
In addition, it is also an important issue for handling and storage of this chemically active and radioactive material~\cite{Cunningham1993,Haschke2001}. 
Of considerable concern is the corrosion products which are formed by reaction of plutonium with water, 
air and hydrogen~\cite{Hodges1979b,Hodges1979,Haschke1983,Haschke1984,Haschke1992,Haschke1995,Allen1998,Haschke1998,Haschke2001}. 
These corrosion products, such as Pu$_2$O$_3$ and PuH$_x$, were suggested to promote the corrosion of metal.

Plutonium monoxide monohydride (PuOH) is a potentially reactive compound formed by corrosion of Pu in liquid water or moisture at room temperature~\cite{Haschke1983,Haschke1984,Haschke1992,Haschke1995}. 
In particular, PuOH is considered to initiate the rapid hydride-catalyzed corrosion of PuOH-coated plutonium metals~\cite{Allen1998,Haschke2001}, 
and identified as a key product to understand the chloride-catalyzed corrosion of plutonium in glovebox atmospheres~\cite{Haschke1998}. 
The following equation
\begin{eqnarray}
\mbox{Pu (s) + H}_2\mbox{O (l) }\rightarrow\mbox{ PuOH (s) + }\textstyle{\frac{1}{2}} \mbox{H}_2\mbox{ (g)}
\end{eqnarray}
describes the formation of PuOH in near-neutral water 
and also determines the composition of the product from experimental data 
showing that 0.5 mol of H$_2$ is formed by the reaction of 1.0 mol of Pu. 
The identification of PuOH is also based on the thermal gravity (TG), X-ray photoelectron spectroscopy (XPS) and X-ray diffraction (XRD) analyses~\cite{Haschke1984}. 
From the XRD analysis, PuOH has a CaF$_2$-related structure 
formed by a cationic sublattice of Pu (III) and an anionic 
sublattice of O$^{2-}$ and H$^-$.
However, the specific arrangement of O and H atoms is unknown,
which may be ascribed to the small scattering cross section of these light atoms with respect to the heavy Pu atom.
This structural uncertainties have impeded in-depth understanding and further exploration of PuOH.

Moreover, there are very limited experimental reports of the basic properties of PuOH,
which may be attributed to the difficulties in preparing and handling samples.
Besides the above-mentioned partially-determined structure, 
only the thermodynamic properties about the formation enthalpies of PuOH, the reactions of PuOH and PuOH-coated Pu with O$_2$ are estimated~\cite{Allen1998}.
In fact, density-functional study has already been performed to 
deeply understand the basic properties of plutonium metal, 
oxides and hydrides~\cite{Shick2005,Sun2008,Jomard2008,Zhang2010,Sun2012,Ao2012,Ao2012b,Ao2013,Yong2013,Zheng2014,Yang2015}.
Therefore it is interesting and necessary to perform the density functional theory (DFT) calculations 
to systematically investigate the ground state properties of PuOH, including the prediction of the Wyckoff positions of O and H.

%In addition, Ref.~\cite{Allen1998} has already estimated
%the thermodynamic properties of PuOH, 
%which are important to understand the energetics and thermodynamic stability of PuOH, 
%and the reactions of PuOH and PuOH-coated Pu with O$_2$.
%Since there is no theoretical counterpart for these experimental study,
%it is crucial to calculate the ground state properties of PuOH by performing the DFT study.

In this work, we have predicted the ground state crystal structure from the perspective of the energetic stability, 
and systematically calculate the electronic, optical, mechanical and thermodynamic properties of PuOH.
The rest of this paper is organized as follows. 
The computational details are described in section~\ref{sec:detail},
and the results are presented and discussed in section~\ref{sec:result}.
Section~\ref{sec:summary} presents a brief summary.

\section{Computational details}
\label{sec:detail}

This study was performed using the VASP code~\cite{Kresse1996}.
The projector augmented wave (PAW)~\cite{Blochl1994,Kresse1999}
pseudopotentials for Pu, O, and H with the 
$6s^26p^26d^25f^4$, $2s^22p^2$, and $1s^1$ valence
electronic configurations are chosen.
For the exchange and correlation energy,
both the local density approximation (LDA)~\cite{Perdew1981} and the generalized gradient approximation (GGA)
using Perdew-Becke-Ernzerhof (PBE)~\cite{Perdew1996} functional are employed.
In addition, we also studied three possibilities for the magnetic states: nonmagnetic (NM), antiferromagnetic (AFM), 
ferromagnetic (FM).
The kinetic energy cutoff is taken as 700 eV and an $8\times8\times8$
Monkhorst-Pack $k$ point sampling mesh~\cite{Monkhorst1976} of Brillouin zone is used.

In order to capture the localization effect of the $f$ electrons
coming from the strong electron-electron interaction,
we employed the so-called DFT+$U$ approach which has been successfully applied to $\delta$-plutonium~\cite{Shick2005}, 
plutonium oxides~\cite{Sun2008,Jomard2008,Zhang2010,Sun2012} and hydrides~\cite{Ao2012,Ao2012b,Ao2013,Yong2013,Zheng2014,Yang2015}.
Here we choose the rotationally invariant DFT+$U$ scheme introduced by Liechtenstein {\it{et. al.}}~\cite{Liechtenstein1995},
which was already adopt in the study of Pu~\cite{Shick2005} and PuO$_2$~\cite{Jomard2008}.
Unless otherwise stated, the typical set of parameters: $U$=4.0 eV and $J$=0.7 eV, are chosen in the following.
%The problem of metastable states in the DFT+$U$~\cite{Jomard2008} are circumvented by adiabatically switching on the Hubbard $U$ parameter~\cite{Geng2010}.

%Methfessel and Paxton’s smearing method~\cite{Methfessel1989}
%of the first order is used with a
%width of 0.1 eV to determine the partial occupancies for each
%Kohn-Sham orbitals.

%Relaxations are performed using the conjugate gradient and quasi-Newton
%algorithm with a convergence criterion of 1 meV with
%regards to the total free energy of the system.

\section{Results and discussion}
\label{sec:result}

\subsection{Crystal structure}

\begin{figure}[tb]
\begin{center}
\includegraphics[width=0.8\textwidth]{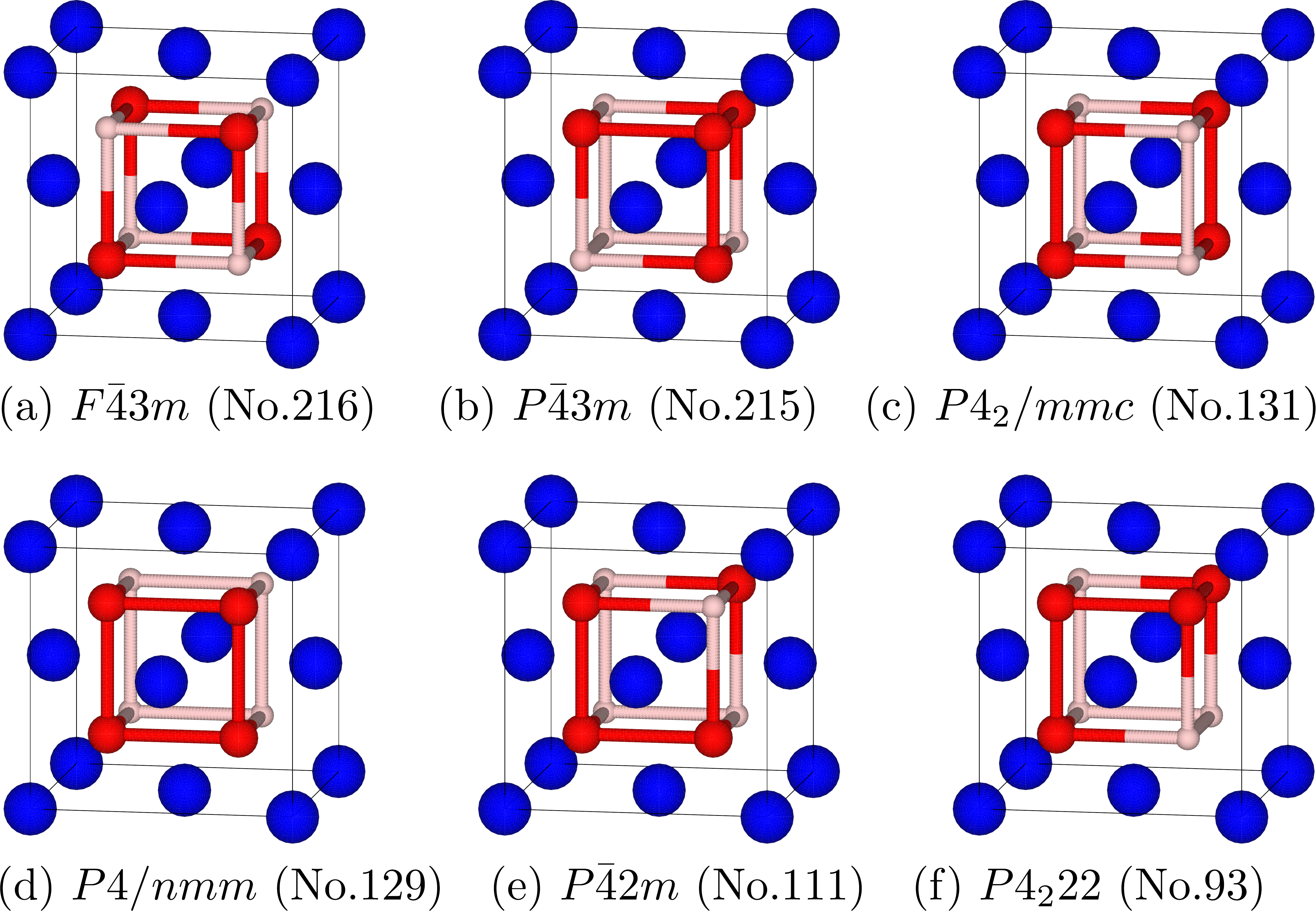}
\caption{Six possible fluorite-related crystal structures of PuOH with the blue, red, pink balls representing Pu, O, H atoms, respectively.}
\label{fig:structure}
\end{center}
\end{figure}

As concerns the fluorite-related structures of PuOH, Pu occupied the positions in an face-centered cubic (fcc) configuration while O or H occupied
the eight tetrahedral sites of the fcc lattice. 
In order to consider all the possible arrangement, 
one can randomly select four tetrahedral sites to accommodate the H atoms and the residual tetrahedral sites for the O atoms.
After symmetry analyses, one can found that there are only six inequivalent atomic configurations, which are shown in Figure~\ref{fig:structure}.
The space groups include $F\bar{4}3m$, $P\bar{4}3m$, $P4_2/mmc$, $P4/nmm$, $P\bar{4}2m$ and $P4_222$,
which would be used to denote the crystal structures below.

For $F\bar{4}3m$ phase (Figure~\ref{fig:structure} (a)), four hydrogen atoms are fully separated from each other and the corresponding symmetry is the highest.
For the structure with the second highest symmetry, $P\bar{4}3m$ phase (Figure~\ref{fig:structure} (b)), 
four hydrogen atoms are connected to each other and constitute a trirectangular tetrahedron.
For $P4_2/mmc$ and $P4/nmm$ phase (Figure~\ref{fig:structure} (c) and (d)), 
four hydrogen atoms are located in the \{110\} and \{100\} plane, respectively.
For $P\bar{4}2m$ phase (Figure~\ref{fig:structure} (e)),
three hydrogen atoms constitute a right-angle and the fourth hydrogen atom is located away from them.
For the structure with lowest symmetry, $P4_222$ phase (Figure~\ref{fig:structure} (f)), four hydrogen atoms constitute two right angles.
If the O-H system in the central region of the fcc lattice is viewed as the O-H mixture, the $F\bar{4}3m$ mixture is the most homogeneous and
the segregation of the $P4/nmm$ and $P\bar{4}3m$ mixtures are the most serious.

\begin{figure}[tb]
\begin{center}
\includegraphics[width=0.95\textwidth]{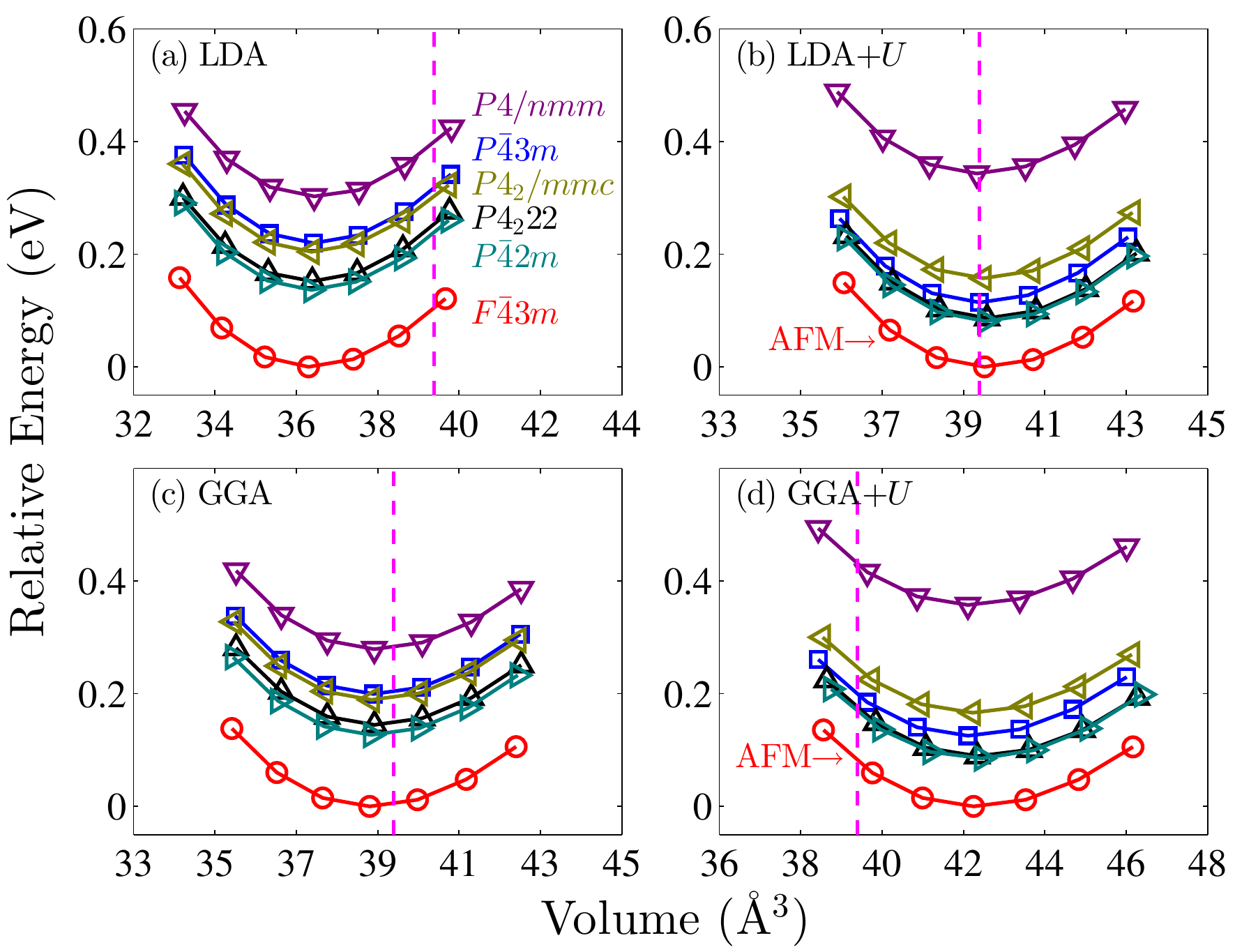}
\caption{Calculated total energies for the six fluorite-related structures of PuOH as a function of volume within 
four exchange-correlation functional approximations: (a) LDA, (b) LDA+$U$, (c) GGA, and (d) GGA+$U$.
The dashed lines denote the experimental volume.
All the magnetic states presented here are FM except AFM $F\bar{4}3m$-phase in the LDA/GGA+$U$ scheme, denoted as the bottom lines in (b) and (d).
}
\label{fig:bulk}
\end{center}
\end{figure}

Total energies for the six possible fluorite-related structures of PuOH are calculated within density functional theory as a function of lattice constant.
For each space group, the total energies are obtained using three magnetic orders (NM, AFM, and FM) and only the lowest one of the three is presented.
In Figure~\ref{fig:bulk}, the total energy for all phases are plotted as a function of volume 
Here and in the following the values of energy and volume are presented per formula unit.
Notice first that our calculation shows that the $F\bar{4}3m$ phase is the most stable whatever the exchange-correlation functional approximation we choose.
Besides, all the relative energies of the other phases are greater than 81 meV (equivalent to 940 K), which implies that PuOH would only adopt
the space group $F\bar{4}3m$ (Figure~\ref{fig:structure} (a)) in the room temperature or intermediate temperature.

Furthermore, the energetics stability ordering of the six fluorite-related structures can also be obtained from Figure~\ref{fig:bulk},
which is 
$F\bar{4}3m$ $<$ $P\bar{4}2m$ $<$ $P4_2/mmc$ $<$ $P4_222$ $<$ $P\bar{4}3m$ $<$ $P4/nmm$ for LDA/GGA,
and 
$F\bar{4}3m$ $<$ $P\bar{4}2m$ $<$ $P4_2/mmc$ $<$ $P\bar{4}3m$ $<$ $P4_222$ $<$ $P4/nmm$ for LDA/GGA+$U$.
Recalling of the above analogy of the structure with the O-H mixture, 
we realized that the system is more stable for lower degree of segregation.
The statement might be universal for various Pu+O+H systems, which signifies that the stability of the system may be achieved
for the homogeneous O-H mixture and hydrogen transport may be rapid for the low-oxygen system.

\subsection{Structural properties}
\label{sec:structure}

\begin{figure}[tb]
\begin{center}
\includegraphics[width=0.7\textwidth]{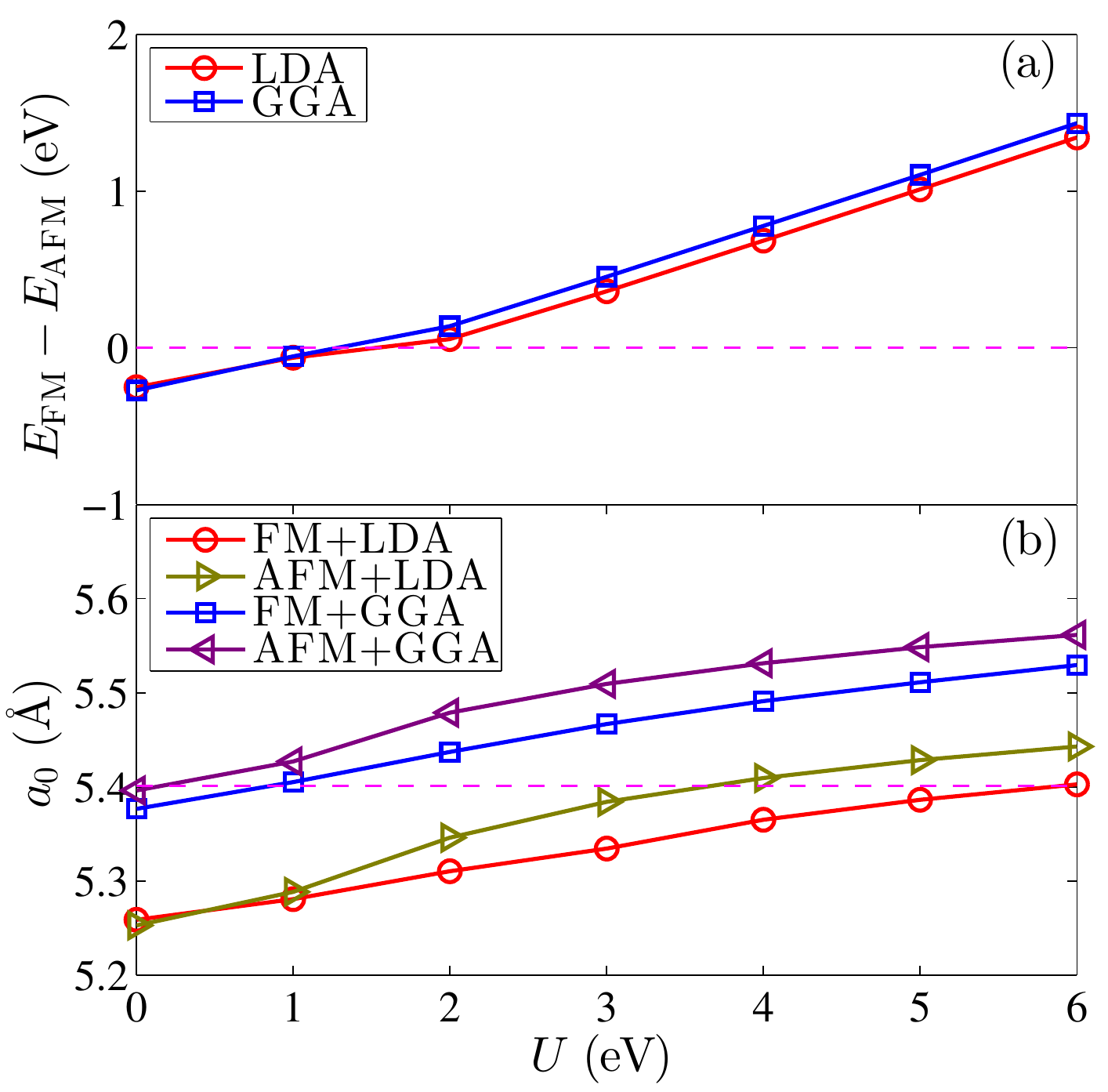}
\caption{Dependence of (a) energy difference between AFM and FM order and (b) equilibrium lattice parameter on Hubbard parameter $U$. 
The dashed line labels the transition and the experimental lattice parameter in (a) and (b), respectively.
Note that $U=0$ means pure LDA/GGA rather than LDA/GGA+$U$ with $U=0$.}
\label{fig:cross}
\end{center}
\end{figure}

Now let us focus on the $F\bar{4}3m$ phase.
In this cubic unit cell defined by lattice parameter $a_0$,
the crystallographic positions are Pu 4$a$ (0,0,0), 
H 4$c$ ($\textstyle\frac{1}{4}$, $\textstyle\frac{1}{4}$, $\textstyle\frac{1}{4}$),
and O 4$d$ ($\textstyle\frac{3}{4}$, $\textstyle\frac{3}{4}$, $\textstyle\frac{3}{4}$) (or H 4$d$ and O 4$c$).
Since the NM phase of PuOH is not energetically favorable 
compared with their FM and AFM phases, 
we will concentrate on the phases with magnetic order.
The magnetic states predicted by our LDA/GGA calculation is FM while that from the LDA/GGA+$U$ calculation is AFM.
To display the FM-AFM transition, 
we show in Figure~\ref{fig:cross} (a)
the dependence of $E_{\rm FM}-E_{\rm AFM}$ on the Hubbard parameter $U$.
For LDA+$U$, the total energy of the AFM phase is lower than of the FM phase for $U$ larger than $\sim$1.4 eV,
while for GGA+$U$, the turning point occurs at a small $U$ of $\sim$1.2 eV.
The total energy differences $E_{\rm AFM}-E_{\rm FM}$ within the LDA+$U$ and GGA+$U$ at $U$=4 eV are 0.68 and 0.78 eV, respectively.
This large value of energy difference suggests that the $F{\bar 4}3m$ phase of PuOH is AFM.

Now let us focus on the lattice parameter $a_0$.
In Figure~\ref{fig:cross} (b) we plot the dependence of $a_0$ on $U$.
The values of $a_0$ are obtained by fitting the energy-volume data with the 
third-order Birch-Murnaghan equation of state (EOS) fitting~\cite{Birch1947}.
The EOS fitting also yields the equilibrium volume $V_0$ and bulk modulus $B$ as follows.
Similar to the study of PuO$_2$ and PuH$_2$~\cite{Sun2008,Jomard2008,Ao2012},
the Hubbard $U$ correction in the LDA/GGA+$U$ approach leads to an increase of the equilibrium lattice parameter.

Both the pure LDA and GGA underestimated the lattice parameter of FM and AFM PuOH, as clearly shown in Figure~\ref{fig:cross} (b).
At a typical value of $U=4$ eV, result of AFM phase within the LDA+$U$ formalism (5.409~\AA) is in close agreement
with the experimental value (5.401~\AA).
Therefore it is suitable for us to focus on the results obtained for $U=4.0$ eV,
which we assumed to be the reference value for PuOH and also chosen in the study of Pu~\cite{Shick2005} and PuO$_2$~\cite{Jomard2008}.

In Table~\ref{tab:structure}, we listed $a_0$, $V_0$, $B$, pressure derivative of the bulk modulus $B'$,
band gap $\triangle$ and magnetic moment of Pu atom $\mu_{\rm mag}$
in the framework of LDA/GGA and LDA/GGA+$U$.
Clearly, our calculations predicted that PuOH is an AFM insulator, which is similar to PuO$_2$.

It is interesting to compare the structure properties of PuOH with that of PuO$_2$~\cite{Jomard2008} and PuH$_2$~\cite{Zheng2014} 
with the same fluorite-related structure.
It is intuitive to regard that there is a monotonic evolution along PuO$_2$ $\rightarrow$ PuOH $\rightarrow$ PuH$_2$.
So it is with the bulk modulus $B$ and energy gap $\triangle$.
But for the lattice parameters, the value of PuOH in the LDA/GGA+$U$ formalism is greater than that 
of PuO$_2$ (5.338/5.444~\AA~\cite{Jomard2008}) and PuH$_2$ (5.337/5.418~\AA~\cite{Zheng2014}).
Actually, the experimental values of these compounds have already confirmed this point, although the difference is much smaller.
In order to understanding this non-monotonic evolution, it is necessary to discuss the corresponding electronic structure in section~\ref{sec:electronic}.

\begin{table}[bht]
\begin{center}
\caption{Equilibrium properties of $F\bar{4}3m$-PuOH.
The lattice parameters $a_0$, volume $V_0$, bulk modulus $B$,
and pressure derivative of the bulk modulus $B'$, band gap $\triangle$,
and magnetic moment of Pu atom $\mu_{\rm mag}$ are reported for LDA, GGA, LDA+$U$, and GGA+$U$.
Experimental values are also shown for comparison.}
\label{tab:structure}
\begin{tabular}{cccccc}
\hline 
Properties & LDA & GGA & LDA+$U$ & GGA+$U$ & Exp.\\
\hline
$a_0$ (\AA) & 5.259 & 5.377 & 5.409 & 5.531 & 5.401~\cite{Haschke1984}\\
$V_0$ (\AA$^3$) & 36.35 & 38.85 & 39.57 & 42.31 & 39.39\\
$B_0$ (GPa) & 150 & 123 & 131 & 111 & \\
$B'$ & 4.4 & 4.4 & 4.1 & 4.1 & \\
$\triangle$ (eV) & 0.0 & 0.0 & 1.298 & 1.683 & \\
$\mu_{\rm mag}$ ($\mu_{\rm B}$) & 4.901 & 4.948 & 4.984 & 5.027 & \\
\hline
\end{tabular}
\end{center}
\end{table}

\subsection{Mechanical properties}

The mechanical properties, by which we understand the behavior of materials under external forces,
are of special importance in fabrication processes and applications.
The mechanical properties of a crystal are described by considering this crystal as a homogeneous
continuum medium, i.e., an elastic body, rather than a periodic array of atoms.
For an elastic body, the Hooke's law holds and the corresponding elastic properties are total determined by the elastic tensor,
which is the coefficient that relates the stress to the strain.
In our work, the elastic tensor is also derived from the strain-stress relationship~\cite{LePage2002}
by performing finite distortions of the lattice and also allowing the internal relaxation of the ions.
For the elastic tensor of a cubic crystal, 
there are only three independent elastic constants, i.e., $c_{11}$, $c_{12}$, and $c_{44}$.

For polycrystal, the mechanical properties are evaluated by the elastic moduli,
including bulk modulus $B$, shear modulus $G$, Young's modulus $E$, and Possion's ratio $\nu$, 
which are the measure of the compressibility, rigidity, stiffness, and Possion's effect, respectively.
The polycrystalline elastic moduli could be calculated by employing the Voigt-Reuss-Hill approximation~\cite{Hill1952}.
In this approach, the effective moduli are approximated
by the arithmetic mean of the two well-known bounds for monocrystals according to Voigt~\cite{Voigt1928} and Reuss and Agnew~\cite{Reuss1929}. 
For a cubic system, the mathematical formulation is 
provided in the following equations~\cite{Hill1952}:
\begin{eqnarray}
B &=& \frac{c_{11}+2c_{12}}{3},\\
G &=& \frac{1}{2}\left(\frac{c_{11}-c_{12}+3c_{44}}{5}+
\frac{5c_{44}(c_{11}-c_{12})}{4c_{44}+3(c_{11}-c_{12})}\right),\\
E &=& \frac{9BG}{3B+G},\\
\nu&=& \frac{3B-E}{6B}.
\end{eqnarray}

%$B$=$\textstyle{1}{3}(c_{11}+2c_{12})$, 
%\begin{eqnarray}
%G = \frac{1}{2}\left(\frac{c_{11}-c_{12}+3c_{44}}{5}+
%\frac{5c_{44}(c_{11}-c_{12})}{4c_{44}+3(c_{11}-c_{12})}\right),
%\end{eqnarray}
%$E$=$9BG/(3B+G)$, and $\nu$=$(3B-E)/(6B)$.

In Debye's theory~\cite{Debye1912,Born1954},
the solid was also treated as the elastic body,
and the lattice vibrations are regarded as the elastic waves.
In order to impose a finite number of modes in the solid, 
Debye used a maximum allowed frequency $\omega_{\rm D}$, 
i.e., the frequency of the highest normal mode of vibration.
The corresponding temperature is the so-called Debye temperature, 
$\theta_{\rm D}=h\omega_{\rm D}/k_{\rm B}$
with $h$ and $k_{\rm B}$ being the Planck and Boltzmann constants,
respectively.
Now the Debye temperature $\theta_{\rm D}$ becomes one of the simplest and the most useful parameters to understand
thermodynamics properties of solid. In addition, $\theta_{\rm D}$ is also a measure of the hardness of the solid~\cite{Abrahams1975}.

The same elastic assumptions imply that 
$\theta_{\rm D}$ could be evaluated in terms of the elastic moduli.
The relation between the Debye temperature $\theta_{\rm D}$ with elastic moduli is given by
\begin{eqnarray}
\theta_{\rm D}
=\frac{h}{k_{\rm B}}
\left[\frac{9n}{4\pi V_0}\right]^{1/3} \left[2\left(\frac{\rho}{G}\right)^{3/2}+\left(\frac{3\rho}{3B+4G}\right)^{3/2}\right]^{-1/3},
\end{eqnarray}
in which $n$ is the number of atoms in one formula and $\rho$ is the mass density.

In Table~\ref{tab:elastic} we listed the calculated elastic constants,
elastic moduli, Poisson's ratio, and Debye temperature of $F\bar{4}3m$-PuOH.
Notice that the bulk modulus $B$ deduced from elastic constants 
are very close to that obtained by EOS fitting in Table~\ref{tab:structure}. 
This indicates that our calculations are consistent and reliable.

Let us focus on the mechanical stability first.
For cubic crystals, the stability criteria requires that
\begin{eqnarray}
c_{44}>0, \quad c_{11}>c_{12},\quad c_{11}+2c_{12}>0,
\end{eqnarray}
which implies that the energy change upon any small deformation is positive.
Obviously, the $F\bar{4}3m$ phase of PuOH is mechanically stable.

Moreover, let us note that the Hubbard $U$ correction in the LDA/GGA+$U$ approach leads to a decrease of the elastic constants and elastic moduli.
As was the case of PuO$_2$~\cite{Zhang2010}.
Next let us compare the mechanical properties of PuOH with that of PuO$_2$~\cite{Zhang2010} with the same fluorite-related structure.
Take the results from LDA+$U$ framework as an example ($c_{11}$=319.6 GPa, $c_{12}$=177.8 GPa, $c_{44}$=74.5 GPa for PuO$_2$~\cite{Zhang2010}).
The values of $c_{11}$ and $c_{12}$ of PuOH are up to 50\% smaller than that of PuO$_2$, 
while the value of $c_{44}$ of PuOH is almost the same as that of PuO$_2$.
This signifies that PuOH is sensitive to compress strain and insensitive to shear strain compare to PuO$_2$.
This point is also confirmed by comparing the values of elastic moduli.

\begin{table}[bth]
\begin{center}
\caption{Mechanical properties of $F\bar{4}3m$-PuOH.
The elastic constants ($C_{11}$, $C_{12}$, $C_{44}$), bulk modulus ($B$), Shear modulus ($G$), 
Young's modulus ($E$), Possion's ratio ($\nu$), Debye's temperature ($\theta_{\rm D}$)
are reported for LDA, GGA, LDA+$U$, and GGA+$U$.}
\label{tab:elastic}
\begin{tabular}{ccccccccc}
\hline
Method & $C_{11}$ & $C_{12}$ & $C_{44}$ & $B$ & $G$ & $E$ & $\nu$ & $\theta_{\rm D}$\\
　& (GPa) & (GPa) & (GPa) & (GPa) & (GPa) & (GPa) &  & (K)\\
\hline
LDA     & 221 & 115 & 79 & 150 & 67 & 176 & 0.31 & 344 \\
GGA     & 194 & 88 & 72 & 123 & 64 & 163 & 0.28 &  337\\
LDA+$U$ & 198 & 96 & 77 & 130 & 65 & 168 & 0.28 & 343 \\
GGA+$U$ & 182 & 73 & 64 & 109 & 60 & 152 & 0.27 & 332 \\
\hline
\end{tabular}
\end{center}
\end{table}

\subsection{Electronic structure}
\label{sec:electronic}

\begin{figure}[tb]
\begin{center}
\includegraphics[width=0.8\textwidth]{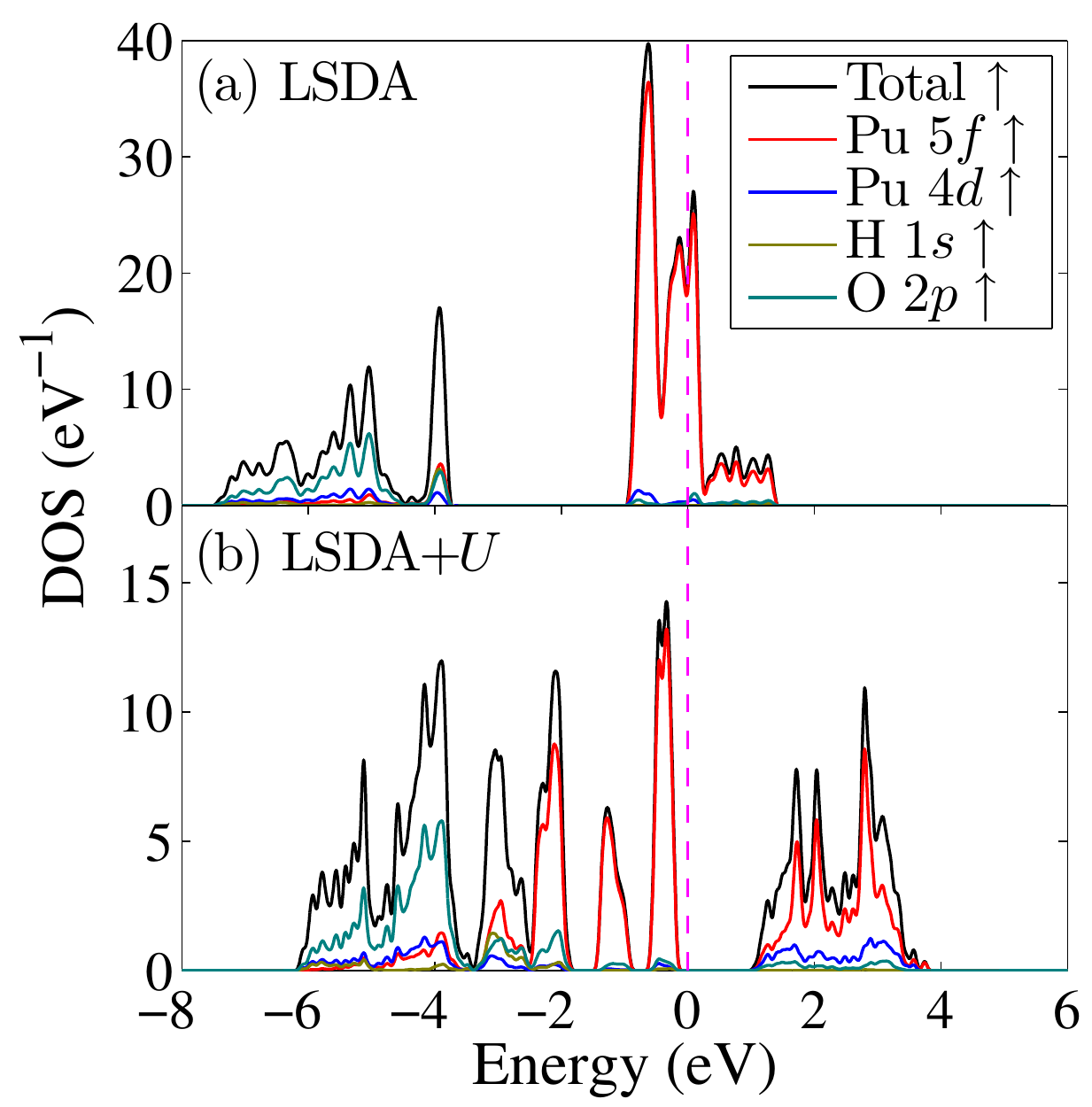}
\caption{The total DOS for the PuOH computed in the (a) LDA, (b) LDA+ $U$ ($U$=4) formalisums.
The projected DOSs for the Pu 5$f$, Pu 6$d$, O 2$p$, and H 1$s$ orbitals are also shown. 
The Fermi level is set at zero and labeled by the dashed line.}
\label{fig:dos}
\end{center}
\end{figure}

The vast array of electrical, optical and magnetic properties of the material are determined by the electronic structure.
In Figure~\ref{fig:dos} we plot the total density of states (DOS) as well as the projected DOS for the Pu 5$f$, Pu 6$d$ O 2$p$, H 1$s$ orbitals of PuOH within
the LDA and LDA+$U$ formalisms. Results from GGA and GGA+$U$ are similar and hence are not shown here.
Obviously, The ground state of $F{\bar 4}3m$-PuOH is predicted to be metallic for the pure LDA and insulating for LDA+$U$,
which is similar to PuO$_2$~\cite{Sun2008,Jomard2008,Zhang2010}.
Based on the ubiquitous localized nature of 5$f$ electron in Pu compound and
the above discussion in section~\ref{sec:structure},
we conclude that PuOH is an antiferromagnetic insulator with space group $F{\bar 4}3m$.

\begin{figure}[tb]
\begin{center}
\includegraphics[width=0.8\textwidth]{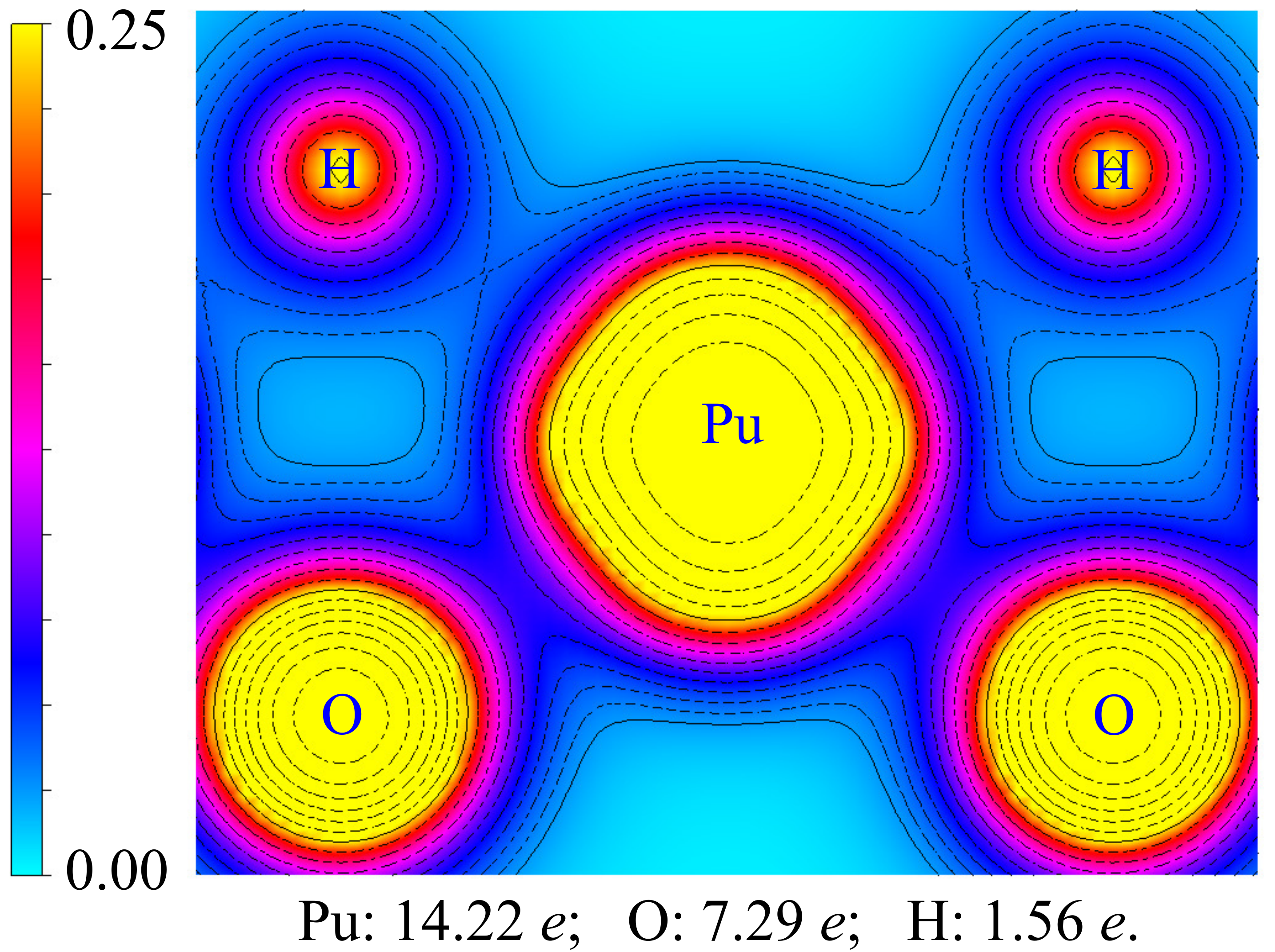}
\caption{A slice of the isosurfaces of the electron density in the (110) plane from the LDA+$U$ calculation.
The bottom list the total valence electron charge for Pu, O, and H atoms according to Bader's partitioning.}
\label{fig:charge}
\end{center}
\end{figure}

From the partial DOS (pDOS), $F{\bar 4}3m$-PuOH is predicted to be a charge-transfer insulator from the pDOS in Figure~\ref{fig:dos}.
This behavior is similar to PuO$_2$~\cite{Prodan2007,Wen2012}, but not Pu$_2$O$_3$ which has the same formal charge with PuOH.
Below the Fermi level, the shape of O-$2p$ pDOS follows that of Pu-$5f$ pDOS, especially for the peak around -2 eV.
These indicate the probable presence of Pu-O covalent bond.
In order to clarify the bonding character in PuOH,
we present in Figure~\ref{fig:charge} the charge-density distribution of (110) plane calculated in LDA+$U$ formalism.
For one Pu atom, there are two nearest-neighboring H atoms and two nearest-neighboring O atoms in one (110) plane, as illustrated in Figure~\ref{fig:charge}.
The charge density distribution around O and H atoms are nearly circular with slight deformation toward Pu atoms.
And the charge density distribution around Pu atom are significantly deformed toward O atoms, 
forming a square-like shape due to the T$_{\rm d}$ point group of 4$a$ Wyckoff position.
The covalent bridges between Pu and O atoms are clear and there is also visible bridges between Pu and H atoms.
Thus we draw the conclusion that the Pu-O bond is stronger than the Pu-H bond in this system.
Since there is a blank area between H and O atoms and their distance is as large as 2.7~\AA~($a_0/2$),
we can deduce that there is no H-O bond in this system and PuOH is not an alkali.
This conclusion agrees with the experimental observation that PuOH is insoluble in near-neutral water~\cite{Haschke1995}.

To understand the ionicity of PuOH, 
we carry out the Bader's charge analysis~\cite{Bader1990} and the results are listed in the bottom of Figure~\ref{fig:dos}.
The charge value of O atom in PuOH is closed to that in PuO$_2$ (7.20 $e$)~\cite{Zhang2010},
and the charge value of H atom is close to that in PuH$_3$ (1.50 $e$)~\cite{Zheng2014}.
These observations manifest that the formal charges of O and H could be regarded as O$^{2-}$ and H$^{-}$, respectively.
In addition, the charge of Pu atom in PuOH is lower than that that in PuO$_2$ (13.60 $e$)~\cite{Zhang2010} but larger than that in PuH$_3$ (14.58 $e$)~\cite{Zheng2014},
which indicates the lower valence state of Pu atom (Pu III).
The results are consistent with the experimental argument~\cite{Haschke1995}.

%In order to see the ionicity of PuOH, results from the Bader's charge analysis are present in Table~\ref{tab:charge}.
%\begin{table}
%\begin{center}
%\caption{Calculated charge and volumes according to Bader partitioning for PuOH.}
%\label{tab:charge}
%\begin{tabular}{ccccccc}
%\hline
%Method & $Q_B$(Pu) & $Q_B(O)$ & $Q_B$(H) & $V_B$(Pu) & $V_B(O)$ & $V_B$(H) \\
% & (e) & (e) & (e)  & (~\AA$^3$)  & (~\AA$^3$) & (~\AA$^3$) \\
%\hline
%LDA     & 14.22 & 7.25 & 1.53 & 18.17 & 11.77 & 6.36 \\
%GGA     & 14.14 & 7.27 & 1.59 & 18.75 & 12.82 & 7.24 \\
%LDA+$U$ & 14.15 & 7.29 & 1.56 & 18.96 & 13.24 & 7.32 \\
%GGA+$U$ & 14.07 & 7.31 & 1.61 & 19.53 & 14.45 & 8.27 \\
%\hline
%\end{tabular}
%\end{center}
%\end{table}

\subsection{Optical properties}

Due to the radioactivity of plutonium compounds,
contactless identification of various materials is very necessary, and obviously one of the useful tools is optical detection.
Thus understanding the optical properties of PuOH is extremely important.
Experimental data are not available now.

\begin{figure}[tb]
\begin{center}
\includegraphics[width=0.8\textwidth]{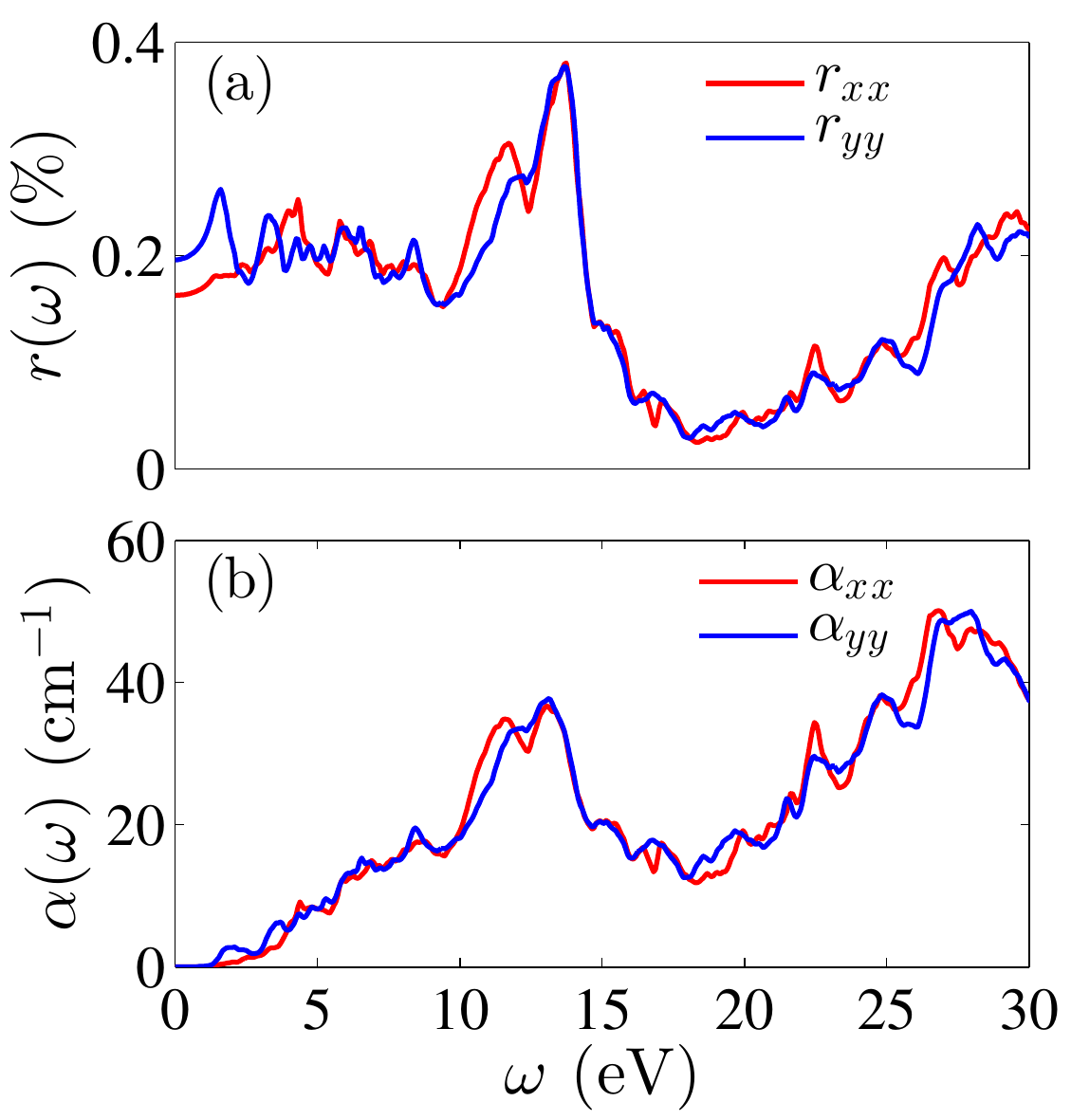}
\caption{(a) The reflectivity and (b) absorption coefficient of PuOH along two directions.}
\label{fig:optics}
\end{center}
\end{figure}

The optical properties as well as the electrical conductivity are determined by the
frequency-dependent dielectric function $\epsilon(\omega)$, 
which has both real and imaginary parts: $\epsilon(\omega)=\epsilon_1(\omega)+i\epsilon_2(\omega)$.
The dielectric function $\epsilon(\omega)$ is determined from the Kubo-Greenwood formula~\cite{Kubo1957,Greenwood1958}
within the independent-particle approximation.
The technical details are described in Ref.~\cite{Gajdos2006}.
The real ($n$) and imaginary ($k$) parts of the index of refraction are related to the dielectric function by a simply formula:
\begin{eqnarray}
\epsilon(\omega)=\epsilon_1(\omega)+i\epsilon_2(\omega)
=\left[n(\omega)+ik(\omega)\right]^2.
\end{eqnarray}
The index of refraction defines the reflectivity ($r$) and absorption coefficient ($\alpha$):
\begin{eqnarray}
r(\omega)=\frac
{[1-n(\omega)]^2+\left[k(\omega)\right]^2}
{[1+n(\omega)]^2+\left[k(\omega)\right]^2},
\quad \alpha(\omega)=2\omega k(\omega).
\end{eqnarray}
The reflectivity and absorption coefficients of PuOH are plotted in Figure~\ref{fig:optics}.
Due to the presence of AFM, the optical properties along the direction of the magnetic moment ($x$ in our study) 
is slightly different from that perpendicular to the direction of magnetic moment ($y$ and $z$ in our study).
The main difference lies in the first peak of $r_{yy}(\omega)$ in Figure~\ref{fig:optics} (a) at 1.6 eV,
which is contributed by the transition from the top of valence bands to the bottom of conduction bands in Figure~\ref{fig:dos}.
This transition also contributes to the rising of the absorption coefficients.
Two main peaks are observed at $11.7$ eV and $13$ eV,
which mainly contributed by the transition to higher energy levels.
Reflectivity of PuOH at $\omega=0$ is not far from the value obtained for PuO$_2$~\cite{Jomard2008,Shi2010}.

%This small energy gap implies that its appearance will be close to that of the metallic plutonium hydride, not the insulator PuO$_2$.
%In fact, experimentally, PuOH is found as as a fine black powder~\cite{Haschke1983}, 
%which is different from the colorful PuO$_2$ but similar to the black hydride.

\subsection{Thermodynamical properties}

All the above discussions are about zero-temperature,
but the state of a physical system is usually relevant to the finite temperature $T$.
From this perspective, it is of great importance to understand the thermodynamical properties of PuOH. 
Usually, the thermodynamical properties are easily measurable and for PuOH, some thermodynamical quantities are already estimated in Ref.~\cite{Allen1998}.

\begin{figure}[tb]
\begin{center}
\includegraphics[width=0.95\textwidth]{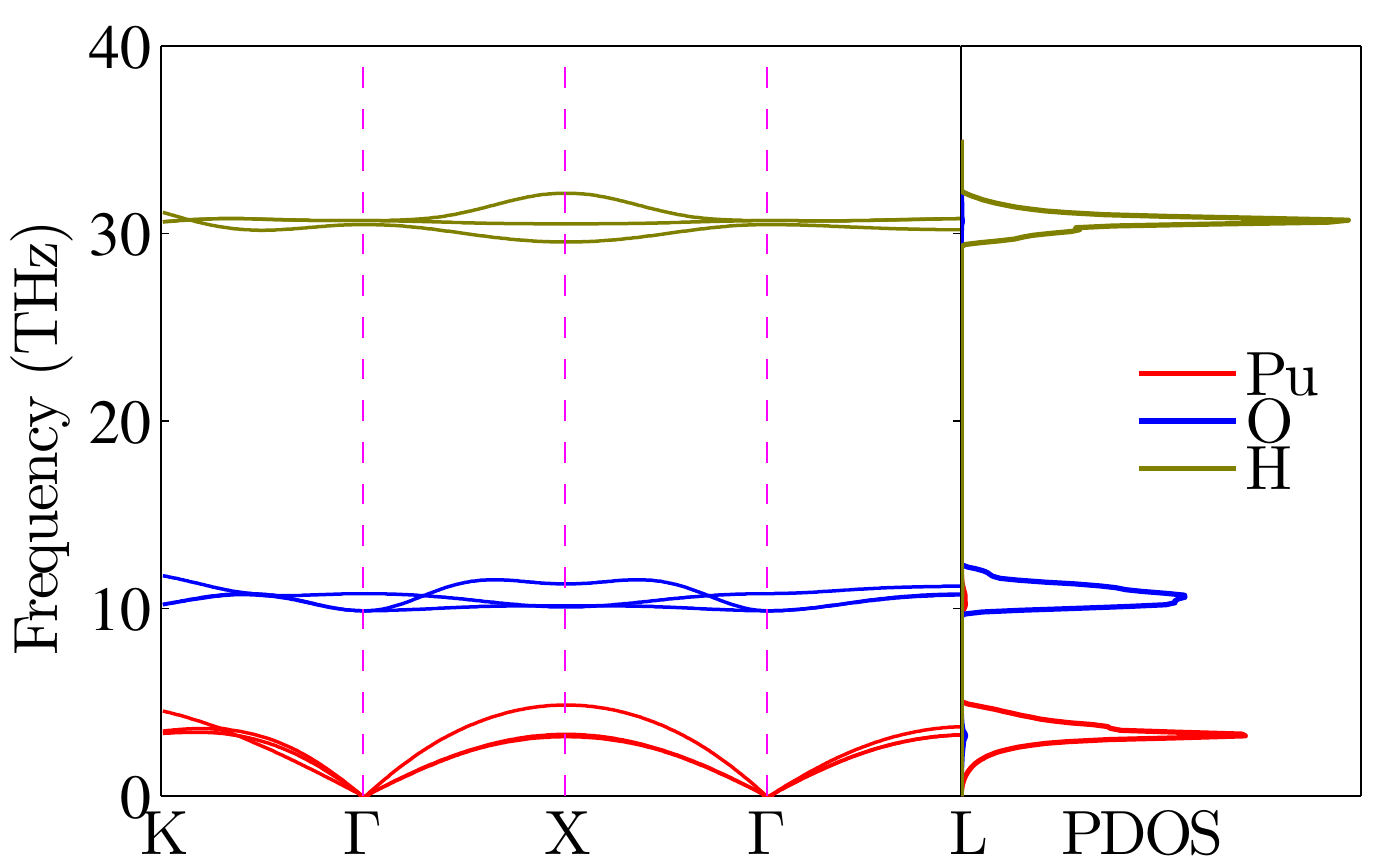}
\caption{Phonon spectrum (left panel) and phonon DOS (right panel) for $F{\bar 4}3m$-PuO$_2$ with the LDA+$U$ fromalisum.}
\label{fig:phonon}
\end{center}
\end{figure}

Thermodynamic properties of crystals are mainly determined by lattice vibrations, i.e., the phonons.
Using the PHONOPY code~\cite{Togo2008}, the phonon frequencies are computed from the real-space force constants.
And the force constants are calculated in the density-functional perturbation theory (DFPT) within the LDA+$U$ scheme.
In Figure~\ref{fig:phonon} 
we plot the phonon spectrum along $K$-$\Gamma$-X-$\Gamma$-L directions in the Brillouin Zone (BZ) and also
the phonon density of states (PDOS).
In the spectrum, there are totally nine phonon modes attributed to the three atoms in the primitive cell.
Due to the mass difference among Pu, O, and H atoms,
Pu atoms contribute the lowest acoustic phonon bands (0-5 THz), 
followed by O atoms (9-12 THz), and finally H atoms (29-32 THz).

\begin{figure}[tb]
\begin{center}
\includegraphics[width=0.8\textwidth]{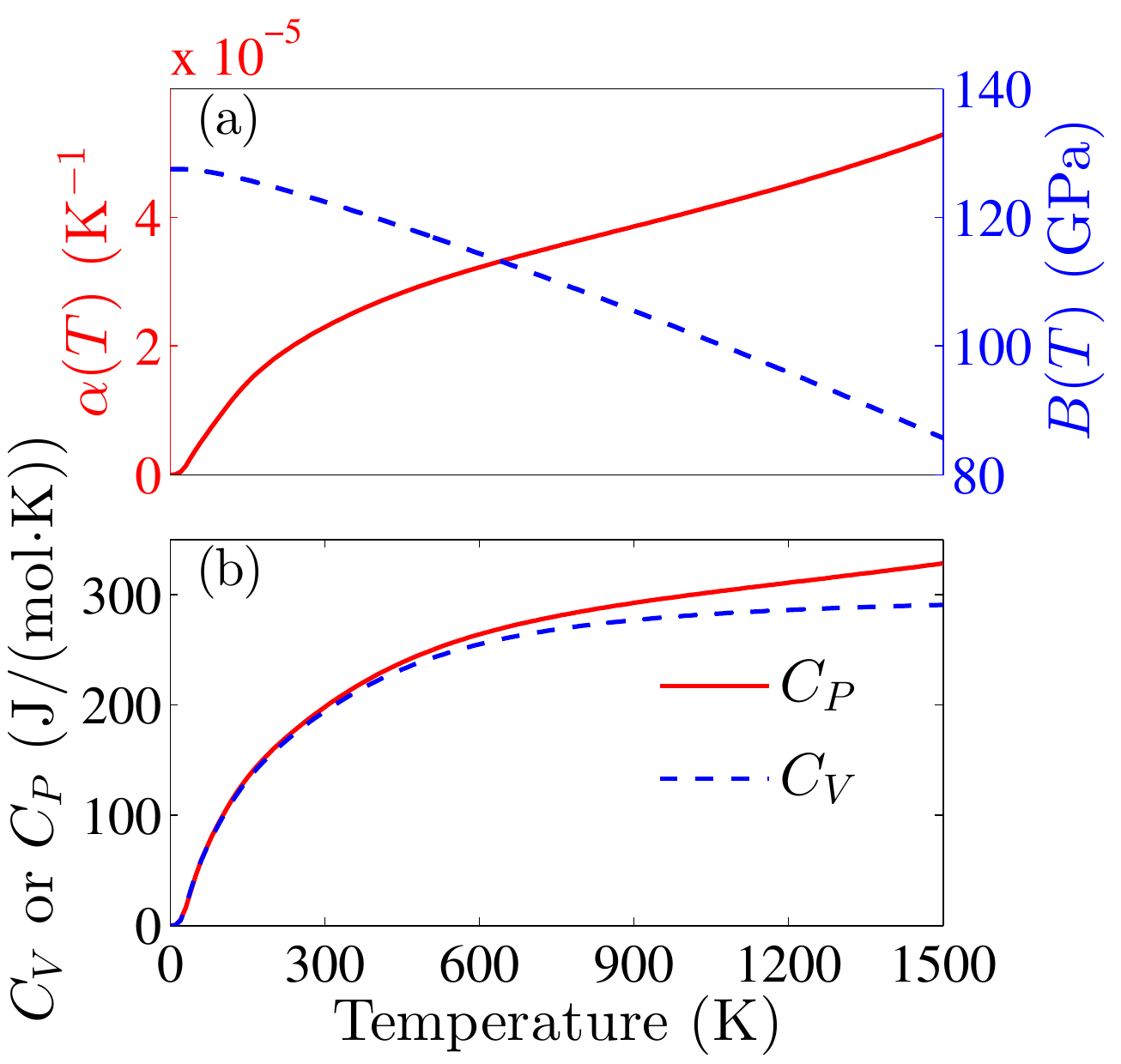}
\caption{(a) The thermal expansion coefficients $\alpha(T)$ and bulk modulus $B(T)$, 
and (b) the specific heat $C_V$ and $C_P$ versus the temperature $T$.}
\label{fig:thermal}
\end{center}
\end{figure}

The thermodynamical quantities at the equilibrium volume $V=V_0$ could be calculated directly by using the above phonon frequencies,
while the determination of the quantities at constant pressure $P$ should be combined within the quasiharmonic approximations
in which the harmonic approximation holds for every value of $V$.
Now The calculated phonon frequency $\omega_{\nu,{\bm q}}(V)$ 
not only depends on the band index $\nu$ and the Bloch wave vector ${\bm q}$,
but also depends on the cell volume $V$.
For specific phonon distribution $\{n_{\nu,{\bm q}}\}$, the lattice vibration
energy is expressed by $E_{\{n_{\nu,{\bm q}}\}}(V)$=$\sum_{\nu,{\bm q}}\hbar \omega_{\nu,{\bm q}}(V)(n_{\nu,{\bm q}}+1/2)$,
and then the corresponding partition function is given by
\begin{eqnarray}
Z(V,T)=\sum_{\{n_{\nu,{\bm q}}\}}\exp\left(-\beta E_{\{n_{\nu,{\bm q}}\}}(V)\right)
=\prod_{\nu,{\bm q}}\frac
{e^{-\beta\hbar\omega_{\nu,{\bm q}}(V)/2}}
{1-e^{-\beta\hbar\omega_{\nu,{\bm q}}(V)}},
\end{eqnarray}
in which $\beta=1/(k_BT)$.
Using this partition function, we could calculate the Helmholtz free energy, $F(V,T)=E(V)-k_BT\ln Z(V,T)$ 
in which $E(V)$ is the internal energy as a function of $V$, 
and the specific heat at constant volume, $C_V=T\left(\frac{\partial^2 F}{\partial T^2}\right)_V$.
Within the quasiharmonic approximation,
the Gibbs free energy as a function of $P$ and $T$ could be determined as
$G(P,T)={\rm min}_V\left[F(V,T)+PV\right]$,
and then the specific heat at constant pressure is given by
\begin{eqnarray}
C_P=T\left(\frac{\partial^2 G}{\partial T^2}\right)_P=C_V+VT\alpha^2(T)B(T),
\end{eqnarray}
where the thermal expansion coefficients $\alpha(T)$ and bulk modulous $B(T)$ are given by
$\alpha(T)=\frac{1}{V}\left(\frac{\partial V}{\partial T}\right)_P$ and 
$B(T)=-V\left(\frac{\partial P}{\partial V}\right)_T$, respectively.

In Figure~\ref{fig:thermal} (a) we plot the thermal expansion coefficients $\alpha(T)$ and bulk modulus $B(T)$ as a function of $T$.
With increasing temperature, the thermal expansion coefficients increase rapidly up to $\sim$300 K, and then become slower.
And the bulk modulus decreases approximately linearly as the temperature is increased.
In Figure~\ref{fig:thermal} (b) we plot the specific heat at constant volume and constant temperature of PuOH as a function of temperature.
At low temperature, $C_V$ equals to $C_P$, while at high temperature,
$C_P$ keeps positive slope while $C_V$ saturates.

At last, we also compute the formation energies $E_{\rm f}^{\rm PuOH}$ with respect to $\delta$-Pu, molecular hydrogen and oxygen, which is written as as follows:
\begin{eqnarray}
E_{\rm f}^{\rm PuOH} = E_{\rm tot}^{\rm PuOH}-E_{\rm tot}^{\rm Pu}-\textstyle{\frac{1}{2}}E_{\rm tot}^{{\rm O}_2}-\textstyle{\frac{1}{2}}E_{\rm tot}^{{\rm H}_2},
\end{eqnarray}
with $E_{\rm tot}^{\rm PuOH}$, $E_{\rm tot}^{\rm Pu}$, $E_{\rm tot}^{{\rm O}_2}$, and $E_{\rm tot}^{{\rm H}_2}$ being 
the total energies of PuOH, $\delta$-Pu, molecular oxygen and hydrogen, respectively.
The energies of reaction for the oxidation of PuOH and PuOH-coated Pu metal,
\begin{eqnarray}
&&
\mbox{PuOH (s) + } \textstyle{\frac{1}{4}} \mbox{O}_2\mbox{ (g) }\rightarrow \textstyle{\frac{1}{2}}\mbox{Pu}_2\mbox{O}_3\mbox{ (s) + }\textstyle{\frac{1}{2}} \mbox{H}_2\mbox{ (g)},
\\
&&
\mbox{PuOH (s) + } \textstyle{\frac{3}{4}} \mbox{O}_2\mbox{ (g) }\rightarrow \mbox{ PuO}_2\mbox{ (s)  + }\textstyle{\frac{1}{2}}  \mbox{H}_2\mbox{O (l)},
\\
&&
\mbox{PuOH (s) + } \textstyle{\frac{1}{4}} \mbox{O}_2\mbox{ (g) + }\textstyle{\frac{1}{3}}\mbox{Pu (s) }\rightarrow
\textstyle{\frac{1}{2}}\mbox{Pu}_2\mbox{O}_3\mbox{ (s) + }\textstyle{\frac{1}{3}} \mbox{PuH}_3\mbox{ (s)},
\\
&&
\mbox{PuOH (s) + } \textstyle{\frac{1}{2}} \mbox{O}_2\mbox{ (g) + }\textstyle{\frac{1}{3}}\mbox{Pu (s) }\rightarrow\mbox{ Pu}\mbox{O}_2\mbox{ (s) + }\textstyle{\frac{1}{3}} \mbox{PuH}_3
\mbox{ (s)},
\end{eqnarray}
are also written as
\begin{eqnarray}
E_{\rm r}^{\rm PuOH\rightarrow Pu_2O_3}&=&
\textstyle{\frac{1}{2}}E_{\rm tot}^{\rm Pu_2O_3}+\textstyle{\frac{1}{2}}E_{\rm tot}^{\rm H_2}-E_{\rm tot}^{\rm PuOH}-\textstyle{\frac{1}{4}}
E_{\rm tot}^{\rm O_2},
\\
E_{\rm r}^{\rm PuOH\rightarrow PuO_2}&=&
E_{\rm tot}^{\rm PuO_2}+\textstyle{\frac{1}{2}}E_{\rm tot}^{\rm H_2}-E_{\rm tot}^{\rm PuOH}-\textstyle{\frac{3}{4}}
E_{\rm tot}^{\rm O_2},
\\
E_{\rm r}^{\rm PuOH+Pu\rightarrow Pu_2O_3}&=&
\textstyle{\frac{1}{2}}E_{\rm tot}^{\rm Pu_2O_3}+\textstyle{\frac{1}{3}}E_{\rm tot}^{\rm PuH_3}-E_{\rm tot}^{\rm PuOH}-\textstyle{\frac{1}{4}}
E_{\rm tot}^{\rm O_2}-\textstyle{\frac{1}{3}}E_{\rm tot}^{\rm Pu},
\\
E_{\rm r}^{\rm PuOH+Pu\rightarrow PuO_2}&=&
E_{\rm tot}^{\rm PuO_2}+\textstyle{\frac{1}{3}}E_{\rm tot}^{\rm PuH_3}-E_{\rm tot}^{\rm PuOH}-\textstyle{\frac{1}{2}}
E_{\rm tot}^{\rm O_2}-\textstyle{\frac{1}{3}}E_{\rm tot}^{\rm Pu}.
\end{eqnarray}
Here $E_{\rm tot}^{\rm Pu_2O_3}$, $E_{\rm tot}^{\rm PuO_2}$ and $E_{\rm tot}^{\rm PuH_3}$
are the total energies of $\alpha$-Pu$_2$O$_3$, PuO$_2$, and fcc-PuH$_3$ compounds, respectively.
In Table~\ref{tab:reaction}, we list these results and also the experimental values for comparison.

\begin{table}[bht]
\begin{center}
\caption{Formation energies of PuOH and reaction energies for the oxidation of PuOH and PuOH+Pu for LDA/GGA and LDA/GGA+$U$.
The experimental values are shown for comparison.
All the energies are in units of eV per PuOH formula unit.}
\label{tab:reaction}
\begin{tabular}{cccccc}
\hline
 & LDA & GGA & LDA+$U$ & GGA+$U$ & Exp.~\cite{Allen1998} \\
\hline
$E_{\rm f}^{\rm PuOH}$
& -6.71 & -6.30 & -7.04 &  -6.34 & -6.55 \\
$E_{\rm r}^{\rm PuOH\rightarrow Pu_2O_3}$
& -1.92 & -1.89 & -2.01 & -1.97 & -2.04\\
$E_{\rm r}^{\rm PuOH\rightarrow PuO_2}$
& -5.79 & -5.05  & -5.69 & -4.90 & -5.46\\
$E_{\rm r}^{\rm PuOH+Pu\rightarrow Pu_2O_3}$
& -2.68 & -2.54 & -2.82 & -2.53 & -2.69\\
$E_{\rm r}^{\rm PuOH+Pu\rightarrow PuO_2}$
& -5.10 & -4.44 & -5.06 & -4.20 & -4.81\\
\hline
\end{tabular}
\end{center}
\end{table}

Within LDA, formation energy deviates from the experiment by 2\% while
for the LDA+$U$ calculation, the difference increase up to 7\%.
For the GGA calculation, $E_{\rm f}^{\rm PuOH}$ deviates from the experiment by 4\%
while within GGA+$U$, the difference decrease to 3\%.
The larger formation energies found
in LDA with respect to GGA are coherent with the overbinding usually found in the LDA approximation. 
For the same reason, the reaction of oxidation of PuOH and PuOH+Pu is more exothermic in LDA(+$U$) than in GGA(+$U$).
All these differences are reasonable and therefore we deduce that our calculation are reliable and consistent.

\section{Summary}
\label{sec:summary}

In this work, we systematically investigate the structural, mechanical, electronic, optical and thermodynamical properties of PuOH
within the framework of density functional theory.
The strong local correlation of $f$ electrons in this compound is taken into account by employing the DFT+$U$ scheme.
Firstly, the crystal structure of PuOH, which is experimentally ambiguous, is identified by the total energy calculation.
PuOH is predicted to have space group $F\bar{4}3m$ (No.~216).
Then the structural properties of $F\bar{4}3m$-PuOH are calculated and the lattice parameters within the LDA+$U$ formalism
is in good agreement with the experimental value.
The mechanical properties such as elastic constants, elastic moduli and Debye's temperature are also computed.

Furthermore, the electronic structures of PuOH are analyzed to reveal the bonding character.
It is found that there is a stronger Pu-O bond and a weaker Pu-H bond.
Bader's charge analysis illustrate the Pu(III) cationic feature, which agrees with the experimental arguments.
The optical properties of PuOH including the reflectivity and absorption coefficient are calculated 
and might be useful to the contactless identification of this compound.

In the last part, the thermodynamical properties of PuOH are presented.
Firstly the phonon spectrum and density of states are given and confirm the dynamical stability of $F\bar{4}3m$-PuOH.
Then some thermodynamical quantities such as the specific heat are presented.
The formation energy of PuOH and the reaction energies for the oxidation of PuOH and PuOH+Pu are calculated.
The results are in reasonable agreement with the experimental values.

\section*{Acknowledgement}

We thank Bo Sun and Yongbin Zhang for useful discussions about $\alpha$-Pu$_2$O$_3$.
This work was supported by
the NSFC (Grant Nos. 11404299, 11305147, and 21471137),
the ITER project (No. 2014GB111006),
the National 863 Program (No. SQ2015AA0100069),
the Science Challenge Program,
the Discipline Development Fund Project (No. ZDXKFZ201206),
the Foundation of President of CAEP (Nos. 2014-1-58 and 2015-2-12),
and the Foundation for Development of
Science and Technology of CAEP (Grant No. 9090707).

\clearpage
\section*{References}
\bibliography{puoh}
\end{document}